\documentclass[a4paper,11pt,english,draft]{article}
\pdfoutput=1

\usepackage{ifdraft}
\ifdraft{\newcommand{\authnote}[3]{{\color{#3} {\bf  #1:} #2}}}{\newcommand{\authnote}[3]{}}

\usepackage[english]{babel}
\usepackage[utf8x]{inputenc}
\usepackage[T1]{fontenc}
\usepackage{fullpage}
\usepackage{verbatim}
\usepackage{filecontents}

\usepackage{titling}
\usepackage[title]{appendix}
\usepackage{framed}
\usepackage{algorithm}
\usepackage{algpseudocode}
\usepackage{color}

\usepackage{mleftright}\mleftright

\usepackage{float}

\let\oldparagraph\paragraph
\renewcommand{\paragraph}[1]{\vspace{-7pt}\oldparagraph{#1}}

 {\endMakeFramed}
\definecolor{mygray}{gray}{0.9}

\usepackage{amsmath}
\usepackage{amssymb}
\usepackage{amsthm}
\usepackage[final]{graphicx}
\usepackage{subcaption}
\usepackage[colorinlistoftodos]{todonotes}
\usepackage[colorlinks=true, allcolors=blue, final]{hyperref}
\usepackage{tikz-cd}
\usepackage{wrapfig}
\usepackage{authblk}

\newcounter{Counter}
\newtheorem{Theorem}[Counter]{Theorem}
\newtheorem{Definition}[Counter]{Definition}
\newtheorem{Lemma}[Counter]{Lemma}
\newtheorem{Corollary}[Counter]{Corollary}

\newcommand{\eps}{\varepsilon}

\newcommand{\nrm}[1]{\left\lVert #1 \right\rVert}
\newcommand{\anrm}[1]{{\left\vert\kern-0.25ex\left\vert\kern-0.25ex\left\vert #1 \right\vert\kern-0.25ex\right\vert\kern-0.25ex\right\vert}}

\DeclareMathOperator{\tr}{tr}
\newcommand{\wt}{\widetilde}

\newcommand{\R}{\mathbb{R}}

\newcommand{\M}{\mathrm{M}}

\definecolor{Tangerine}{rgb}{1,0.5,0}
\definecolor{Fern}{rgb}{0.25,0.5,0}

\title{Finding Angles for Quantum Signal Processing \\ with Machine Precision}
\author[1,4]{Rui Chao}
\author[1,3]{Dawei Ding}
\author[5,6]{Andr\'as Gily\'en}
\author[1,2]{Cupjin Huang\thanks{cupjin.huang@alibaba-inc.com}}
\author[1]{Mario Szegedy}

\affil[1]{\small \textit{Alibaba Quantum Laboratory, Alibaba Group, Bellevue, WA, USA}}
\affil[2]{\small \textit{Department of Computer Science, University of Michigan, Ann Arbor, MI,  USA}}
\affil[3]{\small \textit{Stanford Institute for Theoretical Physics, Stanford University, Stanford, CA, USA}}
\affil[4]{\small \textit{Department of Electrical Engineering, University of Southern California, Los Angeles, CA, USA}}
\affil[5]{\small \textit{QuSoft, CWI, Science Park 123 Amsterdam, The Netherlands}}
\affil[6]{\small \textit{Institute for Quantum Information and Matter, California Institute of Technology, Pasadena, USA}}
\date{}
\begin{document}
\maketitle
\begin{abstract}
We describe an algorithm for finding angle sequences in quantum signal processing, with a novel component we call {\em halving} based on a new algebraic uniqueness theorem, and another we call {\em capitalization}. We present both theoretical and experimental results that demonstrate the performance of the new algorithm. In particular, these two algorithmic ideas allow us to find sequences of more than 3000 angles within 5 minutes for important applications such as Hamiltonian simulation, all in standard double precision arithmetic. This is native to almost all hardware. 
\end{abstract}

\section{Introduction}
Many recent works in quantum computation consider the problem of transforming the spectrum of an unknown unitary in a black box manner.
In a sequence of works~\cite{berry2015HamSimNearlyOpt,childs2015QLinSysExpPrec,apeldoorn2017QSDPSolvers,low2016HamSimQubitization,low2017HamSimUnifAmp,chakraborty2018BlockMatrixPowers,gilyen2018QSingValTransf}
an entire mathematical machinery was developed to address this problem.
In particular, an elegant novel paradigm called \emph{quantum signal processing} was introduced by~\cite{low2016CompositeQuantGates}, which was later organically combined with \emph{qubitization}~\cite{low2016HamSimQubitization,low2017HamSimUnifAmp} in order to address the problem in great generality. 
Quantum signal processing implements polynomial transformations in a very efficient way using only one ancilla qubit with the help of a sequence of single qubit rotation gates. 
In~\cite{gilyen2018QSingValTransf} these techniques were further developed, leading to a Quantum Singular Value Transformation (QSVT) algorithm,
with applications ranging from oblivious amplitude amplification to Gibbs sampling and Hamiltonian simulation. 
However, finding a sequence of angles for the rotation gates corresponding to a desired transformation can be a challenging (classical) task~\cite{childs2017towardsFirstQSimSpeedup}. In order to demonstrate the practicality of quantum algorithms based on quantum signal processing, it is important to show that there is an efficient classical algorithm to find the angles. This question was addressed by~\cite{haah2018ProdDecPerFuncQSignPRoc}, where they showed that this is indeed the case and that the problem can be elegantly treated by considering Laurent polynomials, that is, polynomials with both positive and negative powers. 

\medskip

The main result of our paper can be best summarized 
as a new algorithm for finding these angles, which shows surprising numerical stability. 
Specifically, our paper presents novel contributions to this field in two ways: 
\begin{enumerate}
\item We further develop the mathematics of quantum signal processing by 
defining and an\-a\-lyzing the algebras that naturally arise from the problem, which we identify as Cayley-Dickson algebras~\cite{albert1942quadratic}. In particular, we prove a theorem about the uniqueness of decomposition in the algebra, which is the basis 
for a new algorithm we call {\em halving}.
\item We conduct numerical experiments 
implementing this new algorithm that also make use of a
new method we call {\em capitalization}.
Together, these two algorithmic ideas allow us to use standard double precision arithmetic for finding angle sequences corresponding to high-degree polynomials in quantum signal processing.\footnote{We will make the code of our algorithm open-source via GitHub.}
\end{enumerate}

\medskip

One of the main applications of quantum signal processing is Hamiltonian simulation.
The complexity of simulating Hamiltonians has been an intensively studied topic~\cite{lloyd1996UnivQSim,berry2005EffQAlgSimmSparseHam,berry2015HamSimNearlyOpt,low2016HamSimQSignProc,haah2018QAlgSimLatticeHam}, 
with many results achieving optimal complexity with respect to certain parameters. 
A rich toolbox of algorithmic techniques were developed for simulation including using a linear combination of unitaries~\cite{berry2015HamSimNearlyOpt} and qubitization~\cite{low2016HamSimQubitization}. These techniques proved to be very useful for constructing efficient quantum algorithms for other problems as well.
In this paper we apply our angle finding algorithm to the specific problem of Hamiltonian simulation,  conducting numerical experiments corresponding to evolution time scales two or more orders of magnitude longer than what was previously possible~\cite{childs2017towardsFirstQSimSpeedup}.

\medskip
	
Quantum signal processing can also be applied in settings such as solving linear~\cite{harrow2009QLinSysSolver} and least squares~\cite{wiebe2012QDataFitting} problems with exponential precision, fixed-point amplitude amplification~\cite{yoder2014FixedPointSearch}, robust oblivious amplitude amplification~\cite{berry2015HamSimNearlyOpt}, fast QMA 
amplification~\cite{nagaj2009FastAmpQMA}, fast quantum OR lemma~\cite{brandao2017QSDPSpeedupsLearning}, quantum walk algorithms~\cite{szegedy2004QMarkovChainSearch,magniez2006SearchQuantumWalk}, and quantum machine learning algorithms~\cite{kerenidis2016QRecSys,kerenidis2018QClassSlowFeatAnal}. This can all be done via the QSVT algorithm, which improves upon some of the best known gate complexity results~\cite{gilyen2018QSingValTransf}. Since quantum computing is now entering the noisy intermediate-scale quantum (NISQ)~\cite{preskill2018QuantCompNISQEra} era, it is a key priority to optimize the size of quantum circuits.

\medskip

\emph{Related Concurrent Work:} We initially publicized our results at the 2018 Workshop on Quantum Machine Learning~\cite{szegedy2018ANewAlgorihm} and the 2019 Quantum Information Processing~\cite{szegedy2018FindingAngles} conferences. A related independent work appeared on arXiv~\cite{dong2020efficientPhaseFindingInQSP} which describes a different method (based on black-box optimization) for computing sequences of angles using machine precision arithmetic.

\section{Quantum signal processing}
Quantum signal processing~\cite{low2016CompositeQuantGates} is like a no-look pass in basketball: our goal is to build a quantum circuit that 
transforms a black box operator --- we may call it the {\em signal} --- without peeking
into the box itself. The black box operator is in most cases a quantum circuit itself, so it is unitary.
This will be our assumption throughout.

\medskip

An example of quantum signal processing is when we run a unitary $W$ twice ($C = WW$). The new operator becomes $W^{2}$ regardless of what $W$ is,
and in the process we did not look at $W$.
In this very simple case the Hilbert space, $\mathcal{H}$, of our circuit $C$ was the same as the Hilbert space of $W$ --- no ancilla wires were added. 
In more interesting cases we assume that we have access to a particular controlled version of $W$, which 
we denote by $\widetilde{W}$ and which acts 
on the Hilbert space $\widetilde{\mathcal{H}} = \mathbb{C}^{2}\otimes \mathcal{H}$:
\[
\widetilde{W} =
\left(\begin{array}{ll}
W & 0 \\
0 & W^{-1}
\end{array}\right).
\]
By making use of the extra {\em control wire}  we can 
now design circuits that interleave one-qubit actions on the control wire~\cite{low2016CompositeQuantGates,low2016HamSimQubitization,haah2018ProdDecPerFuncQSignPRoc} with $\widetilde{W}$ operations as follows:
\begin{equation}\label{sigproc}
C = M \cdot Q_{d}  \widetilde{W} Q_{d}^{-1} \cdots
 Q_{1} \widetilde{W} Q_{1}^{-1} \cdot Q_{0} .
\end{equation}
where every $Q_{j}$ is $Q_{j}'\otimes I_{\mathcal{H}}$, and $M$ is a post-selection 
operation:

\paragraph{Post-selection Operator.} Consider a quantum circuit $C = M \cdot C'$ 
with one ancilla qubit and $N$ input wires, which,
setting the ancilla qubit to $|0\rangle$, runs
\begin{equation}\label{repsuc}
 C' = 
\left(\begin{array}{ll}
C'_{00} & C'_{01} \\
C'_{10} & C'_{11}
\end{array}\right) \;\;\;\;\;\;\;\;\;\;\;\; C'_{00}, C'_{01}, C'_{10}, C'_{11} \in \mathrm{M}(2^{N},\mathbb{C})
\end{equation}
and finally performs a post-selection
operation $M$ on the ancilla wire. This latter operation declares success if the ancilla is measured 0 in the computational 
basis and outputs the state in the input wires; otherwise
it runs $C'$ again and again, until success is attained, and only then outputs. A small calculation shows that the thus-obtained action
on the $N$ qubits is not unitary (although $C'$ itself unitary) but is defined by 
the linear operator $C'_{00}$ in the fashion: $|\psi\rangle \rightarrow \frac{C'_{00}|\psi\rangle}{|C'_{00}|\psi\rangle|} $.
The success probability matters: the post-selection process works well if the probability of success,
which is $|C'_{00}|\psi\rangle|$, is not very small.

\medskip

In equation (\ref{sigproc}) we assume that $Q_{j}'$ ($0\le j \le d$) is an arbitrary element of $SU(2)$, the group of two dimensional unitary matrices with unit determinant.
Later we will restrict $Q_{j}'$ to be $X$-rotations, which will still allow us to build most, if not all, of the useful signal processing circuits.

\paragraph{Laurent Polynomials.} Because the blocks of $\widetilde{W}$ are $W$, $W^{-1}$, or zero, 
and because each conjugation by $Q_{j}$ only linearly combines the four blocks
of the operator it receives, the sequence of operations 
in equation (\ref{sigproc}) without the final $M$ represents an operator whose 
all four blocks are Laurent polynomials~\cite{haah2018ProdDecPerFuncQSignPRoc} of $W$. The final $M$ just picks the top left block of this operator
which is itself an operator on the non-ancilla wires
and has the form $F(W) = \sum_{-n}^{n} c_{i} W^{i}$ for some $n$ and $c_{i}$. It is easy to see that $n\le d$. 
From what we have said about post-selection, $C$ acts on $H$ according to $|\psi\rangle \rightarrow \frac{F(W)|\psi\rangle}{|F(W)|\psi\rangle|}$.
The effect of $C$ on an arbitrary operator $W$ is said to be described by $F(w) = \sum_{-n}^{n} c_{i} w^{i}$ if $F(W)$ is the operator to which $C$ transforms $W$.

\medskip

In practice (Hamiltonian simulation, etc.), the Laurent polynomial $F$ is given to us, and our goal is to
design the sequence $Q_{0},Q_{1},\ldots,Q_{d}$, which produces it.
A small but very useful lemma will help us achieve this goal:

\begin{Lemma}
If $C$ has the effect $\sum_{-n}^{n} c_{i} W^{i}$ on all {\em one dimensional} unitary operators $W\in U(1)$,  
then it has the same Laurent polynomial as effect on {\em all} unitary operators.
\end{Lemma}

\noindent For the proof of the lemma let us investigate how $C'$ acts on a state $\underbrace{|\phi\rangle}_{\rm ancilla} \otimes \underbrace{|\psi\rangle}_{\mathcal{H}}$,
where $|\psi\rangle$ is an eigenvector of $W$ with eigenvalue $e^{i\theta}$.
The $Q_{j}$ operations change only the first register and not the second. 
The operator $\widetilde W$ gives a controlled phase shift of the second register, which 
by the ``kick-back'' effect can be represented as a 
$\left(\begin{array}{ll}
e^{i\theta} & 0 \\
0 & e^{-i\theta}
\end{array}\right)$
operation on the first register and no change on the second register. Therefore the subspace $\mathbb{C}^{2} \otimes |\psi\rangle$ 
remains invariant under $C'$ and furthermore, restricted to this subspace, $C'$ acts as if we plugged in 
for $\wt W$ the controlled (i.e. ``tilde'') version of the one dimensional unitary $(e^{i\theta})$. This is already sufficient 
to argue that $W$ is transformed in the same way as the one dimensional unitaries, since all its 
eigenvectors uniformly do so. 

We can in fact explicitly give $F(w) = \sum_{-n}^{n} c_{i} w^{i}$. Let $w$ denote $e^{i\theta}$. Then, when we plug in the one dimensional $w$ into $C'$, we get:
\begin{eqnarray} \nonumber
C'(w) = 
\left(\begin{array}{ll}
C'_{00}(w) & C'_{01}(w) \\
C'_{10}(w) & C'_{11}(w)
\end{array}\right) = 
\hspace{1.6in} \\[8pt]\label{qubit1}
Q_{d}' \cdot 
\left(\begin{array}{ll}
w & 0 \\
0 & w^{-1}
\end{array}\right)
\cdot {Q}_{d}^{\prime -1} 
\cdots
\cdot Q_{1}' \cdot 
\left(\begin{array}{ll}
w & 0 \\
0 & w^{-1}
\end{array}\right)
\cdot {Q}_{1}^{\prime -1}
\cdot Q_{0}'.
\end{eqnarray}
Here $C'_{00}(w),\ldots, C'_{11}(w)$ are Laurent polynomials of $w$. Finally, with $M$ we get $F(w) = C'_{00}(w)$.

\paragraph{\bf An algebra emerges.} The design goal is now clear: Find 1-qubit unitaries $Q'_{0},\ldots,Q'_{n}$ and Laurent polynomials $C'_{01}(w)$,
$C'_{10}(w)$, $C'_{11}(w)$ such that equation (\ref{qubit1}) holds with $F(w) = C'_{00}(w)$.
Later we will see that it is sufficient to restrict ourselves to $d \le n$. Further, we also have the restriction that 
$C'$ is unitary for all unit complex numbers $w$. 

\medskip

Let
M(2, $\mathbb{C}[w,w^{-1}]$) denote the
ring of 2 by 2 matrices over the Laurent polynomials with complex coefficients. 
We may view this ring as an algebra over the real or over the complex numbers.
It turns out that the most natural view of the
algebras that arise in our investigation is to treat them as algebras {\em over the real numbers}.
This algebra certainly contains all 
operators that, {\em when we treat $w$ as a single unknown}, 
we may ever encounter in the expression~\eqref{qubit1} as a partial product. 

\medskip

A way to attain our goal is 
then to do the following two steps:
\begin{enumerate}
\item \textbf{Completion}: Find $C'_{01}(w)$, $C'_{10}(w)$, $C'_{11}(w)$ such that these together with $C'_{00}(w) = F(w)$ form the four components
of $C' \in \mathrm{M}(2, \mathbb{C}[w,w^{-1}])$ that is unitary for every unit value of~$w$.
\item \textbf{Decomposition}: Find constant term ${Q'}_{0}$ and terms
$Q_{i}' 
\left(\begin{array}{ll}
w & 0 \\
0 & w^{-1}
\end{array}\right)  {Q'}_{i}^{-1}$. ($1\le i\le d$) linear in $w, w^{-1}$ such that their product gives $C'$.  
That indeed $d=n$ can be taken we will prove later. 
\end{enumerate}

\vspace{-\parskip}
These two steps are the way to go in Haah's paper~\cite{haah2018ProdDecPerFuncQSignPRoc}. We will propose a new algorithm for the second step based on a new theorem as well as a new method for preprocessing $F$, and obtain excellent 
experimental results. The fact that a rich, well-structured subset of elements of M(2, $\mathbb{C}[w,w^{-1}]$)
can be split as a product of degree-one elements was first observed in~\cite{low2016CompositeQuantGates}. We strengthen this result with a uniqueness theorem 
that allows us to do the decomposition in a binary tree manner rather than sequentially.
Our experiments show a substantial increase in numerical stability when we do the decomposition this way.


\section{Algebras associated with quantum signal processing}

In the previous section we have seen how quantum signal processing translates to a task
of 1. \textbf{Completion} and then 2. \textbf{Decomposition} in M(2, $\mathbb{C}[w,w^{-1}]$).
In this section we define two useful sub-algebras of M(2, $\mathbb{C}[w,w^{-1}]$),
where both 1. and 2. are already possible and the number of free parameters is reduced. We would like to remind the reader that we view M(2, $\mathbb{C}[w,w^{-1}]$)
as an infinite-dimensional algebra {\em over the real numbers}.

\paragraph{The Low algebra.} To define the smaller of the two sub-algebras of M(2, $\mathbb{C}[w,w^{-1}]$) 
(and for practical purposes the more interesting one),
we write down a general element of the algebra in terms of the variable 
\[
\widetilde{w} = \left(\begin{array}{ll}
w & 0 \\
0 & w^{-1}
\end{array}\right).
\]

\vspace{-\parskip}
The motivation comes from the paper of Low et al.~\cite{low2016CompositeQuantGates} that implicitly worked in this algebra.
Let $I, X, Y, Z$ be the Pauli matrices as usual. Then already 
expressions of the form
\begin{equation}\label{exp1}
A(\widetilde{w}) + B(\widetilde{w})\cdot iX  \;\;\;\;\;\;\; A,B \in \mathbb{R}[w,w^{-1}]
\end{equation}
contain all operators that {\em when we treat $w$ as a single unknown} we may ever encounter as a partial product
when $Q_{0}',\ldots,Q_{d}'$ are all $X$-rotations:
\[
\widetilde{\alpha} = \;
\left(\begin{array}{ll}
\cos\alpha & i\sin\alpha \\
i\sin\alpha & \cos\alpha
\end{array}\right)
= \cos\alpha\cdot I + \sin\alpha\cdot iX.
\]

\vspace{-\parskip}
Low et al.\ has shown that this is all we need when our target $F$ has real coefficients and satisfies a parity constraint.
Furthermore, if we change the post-selection by 
introducing a more general final measurement, this framework encapsulates a very rich set of $F$'s, so we need not go further.

\paragraph{The Haah algebra.} If we nevertheless want to obtain all possible Laurent polynomials in the upper left corner including those with complex coefficients,
there is a bigger sub-algebra~\cite{haah2018ProdDecPerFuncQSignPRoc} of M(2, $\mathbb{C}[w,w^{-1}]$) that accomplishes that.
The elements of this sub-algebra are written as
\begin{equation}\label{exp2}
A(\widetilde{w}) + B(\widetilde{w})\cdot iX + C(\widetilde{w})\cdot iY + D(\widetilde{w})\cdot iZ  \;\;\;\;\;\;\; A,B,C,D \in \mathbb{R}[w,w^{-1}].
\end{equation}

\begin{figure}
    \centering
    \includegraphics[width=1.0\textwidth]{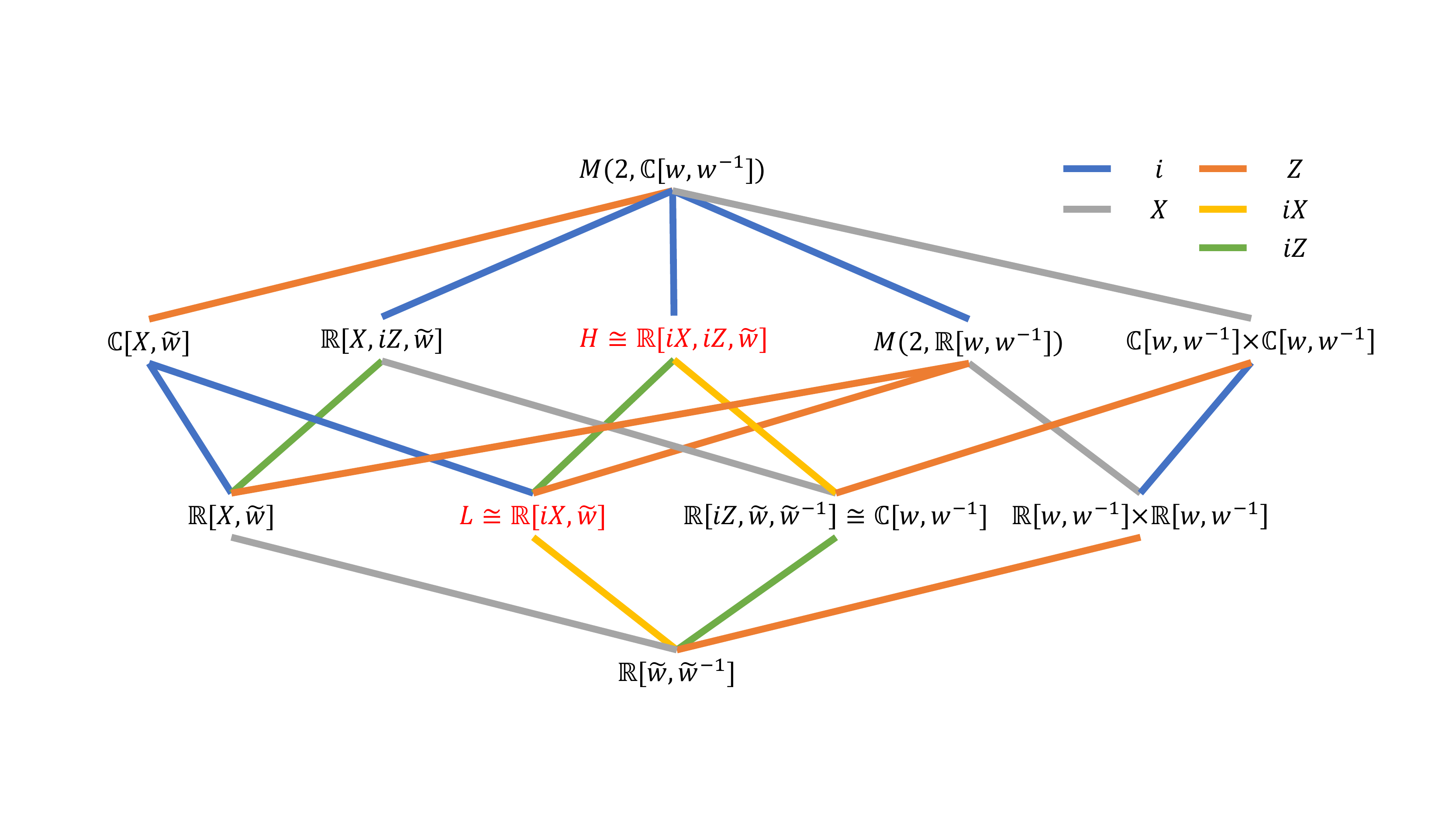}
    \caption{A diagrammatic illustration of the relationship of various Laurent polynomial algebras, where the algebras are identified under $\mathbb{R}$-algebra homomorphisms. The Low and Haah algebras are denoted as $L$ and $H$ and are marked red. Each line indicates a possible generator to add in the algebra below to generate the algebra above. Note that such choices might not be unique.}
    \label{fig:diagram}
\end{figure}

Figure~\ref{fig:diagram} shows a diagram of the relationship of multiple Laurent polynomial algebras.

\section{The syntactic versus semantic view}
Let us make here some clarifying remarks about our expressions. 
When writing down polynomials, one  way of viewing them (Laurent or otherwise) is {\em semantic}. In this interpretation polynomials 
are functions from a set (often a field $\mathbb{F}$, vector space, algebraic variety) to a ring, module, etc.
Another way of viewing polynomials is {\em syntactic}. In this case two polynomials are different if the
sequence of their coefficients differ (formal polynomials).

\medskip

In general, the two views can lead to different definitions. Consider for instance $x^{3}$ in GF(2). By Fermat's little theorem
(or simply by checking both replacements in GF(2)) we conclude that semantically $x^{3}$ is the same as $x$. 
Syntactically however they are obviously different.

\medskip

In our case the domain over which we should view expressions (\ref{exp1}) and (\ref{exp2}) is 
the unit circle $U(1) = \{w \mid \lvert w\rvert = 1\}$. This is because when we instantiate $w$, it is a
one dimensional unitary operator, that is a phase. Now, $\widetilde{w}$ is an element of 
M(2, $\mathbb{C}[w,w^{-1}]$), giving matrix values to expressions (\ref{exp1}) and (\ref{exp2})
with Laurent polynomials as elements. 
We may then ask whether it can occur that two expressions 
in (\ref{exp2}) do not equal syntactically, but they equal semantically.
It is easy to show however, using the fundamental theorem of algebra, that this may not happen.
This permits us to switch back and forth between the two 
interpretations as suits us.

\section{Star operation, unitary and Hermitian elements, and degree}

To define the star operation in M(2, $\mathbb{C}[w,w^{-1}]$) 
we take the semantic view: the star, $M^{\ast}$ of an element 
$M \in\; $M(2, $\mathbb{C}[w,w^{-1}]$)
will have the property that if we 
instantiate $M$ and $M^{\ast}$ over any given
$w\in U(1)$, then we 
get two matrices that are conjugate 
transposes of each other. 
Thus e.g. $\widetilde{w}^{\ast} = \widetilde{w}^{-1}$,
because $w$ and $w^{-1}$ 
are conjugates of each other for any $w\in U(1)$.
On constant matrices (i.e. without the variable $w$)
the star operator works as usual:
\[
(iX)^{\ast} = -iX \;\;\;, \;\;\;
(iY)^{\ast} = -iY \;\;\; , \;\;\; 
(iZ)^{\ast} = -iZ.
\]
In general, if 
$a,b,c,d \in \mathbb{C}[w,w^{-1}]$
and we obtain $a^{\ast}$ from $a$ by 
conjugating the coefficients and swapping 
$w$ and $w^{-1}$ (similarly for $b$, $c$, $d$), then we can define the star operation and \emph{unitarity} for an element in $\mathrm{M}(2, \mathbb{C}[w, w^{-1}])$ as
\[
\left(\begin{array}{ll}
a & b \\
c & d
\end{array}\right)^{\!\!\ast} 
=
\left(\begin{array}{ll}
a^{\ast} & c^{\ast} \\
b^{\ast} & d^{\ast}
\end{array}\right) \; ;
\;\;\;\;\;\;\;\;\;\;\;
M\;\mbox{is unitary} \; \leftrightarrow \; MM^{\ast} = I.
\]
This definition leads to:

\begin{Lemma}
The Haah and Low algebras are closed under the star operation of {\rm M(2, $\mathbb{C}[w,w^{-1}]$)}. (In fact, they are both Cayley-Dickson algebras.)
\end{Lemma}

\vspace{-6pt}
\paragraph{Hermitian elements.} Define a \emph{Hermitian} element $x$ as one that satisfies $x^* = x$ and \emph{anti-Hermitian} as $x^* = -x$. Let us now determine such elements of the Haah algebra.
Notice that the following elements of the Haah algebra are Hermitian:
\begin{equation}\label{herm}
\widetilde{w}^{j} + \widetilde{w}^{-j}\;\;\;\;\;,\;\;\;\;\; (\widetilde{w}^{j} - \widetilde{w}^{-j})\cdot iZ \;\;\;\;\;\;\;\; j \in \mathbb{N}
\end{equation}
and therefore all their linear combinations. 

\begin{Lemma}\label{hermitians}
The set of the Hermitian elements of the Haah algebra consists exactly of linear combinations 
of $\widetilde{w}^{j} + \widetilde{w}^{-j}$,  $(\widetilde{w}^{j} - \widetilde{w}^{-j})\cdot iZ$ $(n\in \mathbb{N})$.
\end{Lemma}
\vspace{-10pt}
\begin{proof}
First notice that the following elements and therefore all their linear combinations are anti-Hermitian, i.e. $M^{\ast} = - M$:
\begin{equation}\label{antih}
\widetilde{w}^{j} - \widetilde{w}^{-j}\;\;\;\;\; ,\;\; \;\;\;  \widetilde{w}^{\pm j}\cdot iX\;\;\;\;\;,\;\; \;\;\;   \widetilde{w}^{\pm j}\cdot iY\;\;\;\;\;,\;\; \;\;\;   (\widetilde{w}^{j}+ \widetilde{w}^{-j}) 
\cdot iZ\;\;\; \;\;\;    \;\;\;\;\;\;\;\;\; j\in \mathbb{N}.
\end{equation}
Next notice that the elements in expressions (\ref{herm}) and (\ref{antih}) together span the Haah algebra.
This finishes the proof, since if a Hermitian element $M$ is written as $M_{1} + M_{2}$, where $M_{1}$ is Hermitian and 
$M_{2}$ is anti-Hermitian, then $M_{2} = M-M_{1}$ must be both Hermitian and anti-Hermitian, so $M_{2}$ must be zero.
\end{proof}
We remark that the set of all Hermitian elements of the Haah algebra form an algebra, which is also the center of the Haah algebra 
(since both $iX$ and $iZ$ are in the Haah algebra, the center can only contain elements that are proportional to $I$).

\paragraph{Degree.} Each element $M(w)$ of $\M(2, \mathbb{C}[w,w^{-1}]$ has a {\em degree}, which is the 
maximum absolute value of any exponent of $w$ that ever occurs in $M$.

\begin{Lemma}
In equation (\ref{exp2}) the maximum degree among $A$, $B$, $C$ and $D$ as formal
Laurent polynomials coincides with the above notion of degree.
\end{Lemma}

When we multiply regular polynomials the degrees add up. The same does not hold for Laurent polynomials, especially ones with matrix coefficients.
The next lemma, crucial for our main lemma, gives a case in the Haah algebra when
we can be certain that the degrees do add up.

\begin{Lemma}\label{hermlemma}
Let $M$ be a Hermitian element of the Haah algebra and $M'$ be an arbitrary element of $\M(2, \mathbb{C}[w,w^{-1}])$. Then $\deg MM' = \deg M + \deg M'$.
\end{Lemma}
\vspace{-10pt}
\begin{proof}
Let $M$ be of degree $n$. By Lemma  \ref{hermitians} we can assume 
that $M = \sum_{j=0}^{n}\lambda_{j} (\widetilde{w}^{j} + \widetilde{w}^{-j}) + \mu_{j} (\widetilde{w}^{j} - \widetilde{w}^{-j}) \cdot iZ$ 
where $\lambda_{j}, \mu_{j} \in \mathbb{R}$ and $\lambda_{n}\pm \mu_{n} i \neq 0$. Let 
\[
M' = 
\left(\begin{array}{ll}
A(w) & B(w) \\
C(w) & D(w) 
\end{array}\right)
\;\;\;\;\in \;\;\;\;   {\rm  M(2, } \mathbb{C}[w,w^{-1}] {\rm )}
\]
be of degree $n'$. We show that the degree of $MM'$ is $n+n'$. Clearly, it is sufficient to show that the degree 
of $M'' = \lambda_{n} (\widetilde{w}^{n} + \widetilde{w}^{-n})M' + \mu_{n} (\widetilde{w}^{n} - \widetilde{w}^{-n}) \cdot iZ M'$ is $n+n'$. We can write $M''$ as
\[
\left(\begin{array}{ll}
((\lambda_{n} + \mu_{n} i ) w^n + (\lambda_{n} - \mu_{n} i ) w^{-n}) A(w) &  ((\lambda_{n} + \mu_{n} i ) w^n + (\lambda_{n} - \mu_{n} i) w^{-n})  B(w) \\
((\lambda_{n} - \mu_{n} i) w^n + (\lambda_{n} + \mu_{n} i ) w^{-n}) C(w) & ((\lambda_{n} - \mu_{n} i ) w^n + (\lambda_{n} + \mu_{n} i) w^{-n})  D(w) 
\end{array}\right)
\]
and it is straightforward to see that the highest degree term cannot cancel whether the exponent is positive or negative.
\end{proof}

\section{Parity Subgroups}
Upon closer inspection of equation~\ref{sigproc}, we notice that the algebra elements we can obtain via quantum signal processing must belong to the following set, which is actually a subgroup of our algebra:
\begin{Definition}[Parity subgroup of real Laurent polynomials ]
The \emph{parity subgroup} $P = P_{0}\cup P_{1}\subset \R [w, w^{-1}]$ is defined to be the subgroup of Laurent polynomials (under multiplication) with parity constraint, that is, an element of this group either has even parity:
\[
P_{0} = \left\{p(w)=\sum_{k=-n}^np_kw^k\mid 
\mbox{$p_k=0$ for all odd $k$}\right\}
\]
or odd:
\[
P_{1} = \left\{p(w)=\sum_{k=-n}^np_kw^k\mid 
\mbox{$p_k=0$ for all even $k$}\right\}.
\]
\end{Definition}
Not all unitary elements satisfy the parity constraint. For instance, the element \[(2I+(w+w^{-1})iX+(w-w^{-1})Y)/\sqrt{8}\]
is a counterexample. Now, $P$ is closed under multiplication but not addition. Starting from $P$, we can define 
the parity subgroup of the 
Haah algebra $H_0\cup H_1$, where $$H_\epsilon := \left\{A(\widetilde{w})+B(\widetilde{w})\cdot iX + C(\widetilde{w})\cdot iY+D(\widetilde{w})\cdot iZ| A,B,C,D\in P_{\epsilon}\right\},\;\;\; \epsilon\in\{0,1\}$$ 
and similarly for the Low algebra. 

\medskip

Low et al.~\cite{low2016CompositeQuantGates} proved an interesting unique decomposition theorem about unitary elements in the Low algebra, which is then generalized by Haah~\cite{haah2018ProdDecPerFuncQSignPRoc} as follows:
	\begin{Theorem}[\cite{low2016CompositeQuantGates,gilyen2018QSingValTransf,haah2018ProdDecPerFuncQSignPRoc}]\label{thm:uniqueDecLow}
		For every unitary parity element $U$ in the Haah algebra with degree $d$, there exists, up to sign, a unique decomposition of $U$ into degree-0 unitary elements and $\tilde w$:
\begin{align}\label{eq:unitaryDecomp}
U(w)=
Q'_d\cdot \widetilde{w}\cdot Q_d^{\prime -1} \cdot \dots\cdot 
Q'_1 \cdot\widetilde{w}\cdot Q_1^{\prime -1} \cdot Q'_0 .
\end{align}
Moreover, when $U$ lies in the Low algebra, the above elements all lie in the Low algebra, i.e.
$Q'_{0},Q'_{1},\ldots,Q'_{d}$ are all
$X$-rotations.
\end{Theorem}
    
\section{The main lemma}

Assume we are given a unitary element, $U$, of the Haah algebra to be decomposed 
into a product of linear terms. Instead of solving for one linear term at a time as in~\cite{haah2018ProdDecPerFuncQSignPRoc}, 
we want to decompose $U$ as $U_{1}U_{2}$ where $U_{1}$ and $U_{2}$ have degrees roughly
half of that of $U$. For any element of the Haah algebra it is proven that such decomposition exists, 
but is it unique? Conceivably, it could occur that $U = V_{1}V_{2}$, where $V_{1}$ and $V_{2}$ are not unitary.
And if we find $V_{1}$ and $V_{2}$ instead of $U_{1}$ and $U_{2}$, we cannot 
continue with the decomposition. A consequence of our Main Lemma 
is that this is impossible.

\begin{Lemma}[Main Lemma]\label{main}
Let $M$ be in the Haah algebra, $M'$ in $\M(2, \mathbb{C}[w,w^{-1}])$. Then 
\[
\deg MM' \ge \deg M' - \deg M + \deg M^{\ast}M.
\]
\end{Lemma}
\vspace{-10pt}
\begin{proof}
Since $\deg M = \deg M^{\ast}$ we have that 
$\deg M^{\ast} M M' \le \deg M + \deg M M'$. On the other hand $M^{\ast} M$ is a Hermitian
element of the Haah algebra, so by Lemma  \ref{hermlemma} we have that 
$\deg M^{\ast} M M' = \deg M^{\ast} M + \deg M'$, which completes the proof.
\end{proof}
We can now prove:
\begin{Corollary}\label{unitaryDecomp}
Let $U$ be a degree-$d$ unitary parity element of the Haah algebra. Then the set of equations
\begin{align*}
\deg(V^{\ast}U)&\leq d-l,\\
\deg(V) &\leq l,\\
V(w = 1)&=I,\\
V & \in H_{\,l \,{\rm mod}\, 2}
\end{align*}
has a unique solution $V$ which is a unitary element of the Haah algebra. Moreover, $V$ lies in the Low algebra if $U$ does.
\end{Corollary}
\vspace{-15pt}
\begin{proof}
We only prove the first part of the theorem. 
The second is an easy consequence.
By Theorem~\ref{thm:uniqueDecLow}, $U$ can be decomposed as
$$U=
\underbrace{Q'_d \cdot\widetilde{w}\cdot Q_d^{\prime -1}}
\cdot \dots\cdot 
\underbrace{Q'_1\cdot \widetilde{w}\cdot Q_1^{\prime  -1} }
\cdot Q'_0,$$
where $Q'_0,\dots,Q'_d$ are all unique. Take 
\begin{eqnarray*}
V & = & \underbrace{Q'_{d} \cdot\widetilde{w}\cdot Q_{d}^{\prime -1}}
\cdot \dots\cdot 
\underbrace{Q'_{d-l+1}\cdot \widetilde{w}\cdot Q_{d-l+1}^{\prime -1}} .
\end{eqnarray*}
It is easy to see that $V$ satisfies the set of equations.

\medskip

To prove uniqueness, suppose that there exists another element $V'$ which satisfies the equations. By Lemma~\ref{main},
$$d-l\geq \deg {V'}^{\ast} U\geq \deg U-\deg V'+\deg V'{V'}^{\ast}\ge d-l+\deg V'{V'}^{\ast}.$$
Therefore, $\deg V'{V'}^{\ast} \le 0$, so $V' V'^*$ is a constant function of $w$. However, $V'(1)=I$, $V^{\prime *}(1) = I$, and so $V' V^{\prime *}(1) =I$. As $V' V'^*$ is constant, $V'$ is a parity unitary and therefore must be equal to $V$ by Theorem~\ref{thm:uniqueDecLow}.
\end{proof}
Note that the conditions on the degrees in Corollary~\ref{unitaryDecomp} are inequalities, but the solution we obtain in the end will satisfy them with equality. This is because the degree of a product can be at most the sum of the degrees.

\section{Algorithm}
We now provide an outline of our angle finding algorithm for quantum signal processing. Suppose we wish our circuit to have the effect of $F(w) \in \mathbb{R}[w, w^{-1}]$. We define the norm of a Laurent polynomial as
\begin{align}
\nrm{F} \equiv \max_{w \in U(1)} |F(w)|.
\end{align}
Since we can always multiply $F$ by a constant factor, thereby incurring a constant factor cost in success probability, we can assume $\nrm{F} < 1$. This is sufficient to run  the completion step of the algorithm, as shown in \cite{low2016CompositeQuantGates, gilyen2018QSingValTransf, haah2018ProdDecPerFuncQSignPRoc} and as we will see in the following.

\paragraph{Capitalization}
In many applications of quantum signal processing, $F(w)$ contains terms with very small coefficients, especially for the higher order terms. This can arise for instance when we are trying to achieve the effect of some analytic function and we use a Taylor series approximation. In~\cite{haah2018ProdDecPerFuncQSignPRoc}, this is cited as the primary source of numerical instability. We propose an ad hoc solution we call \emph{capitalization}. Namely, given that we wish to perform quantum signal processing to some error tolerance $\eps$, we can add leading order terms to $F(w)$ with coefficients on the order of $\eps$ and then run our algorithm. Combined with our new algorithm, we show in our experiments that this does extremely well empirically, allowing us to solve instances orders of magnitude larger than what was previously possible, for example in~\cite{childs2017towardsFirstQSimSpeedup}.

\medskip

A similar preprocessing technique was used in~\cite{haah2018ProdDecPerFuncQSignPRoc}, where all the coefficients of the given Laurent polynomial are rounded to multiples of $\epsilon/n$, where $n$ is the degree. Such rounding would be an alternative way to resolve the numerical instability.

\paragraph{Completion via root finding}
The first step is completion, namely finding a unitary $U$ in the Low algebra such that the upper left corner of $U$ is $F(w)$. To do this, we simply solve for the real Laurent polynomial $G$ such that $F(\wt w) + G(\wt w) \cdot iX$ is a unitary element of the Low algebra. It is not difficult to see that this translates to the equation
\begin{align*}
F(w) F(w^{-1}) + G(w) G(w^{-1}) = 1.
\end{align*}
As done in~\cite{low2016CompositeQuantGates, gilyen2018QSingValTransf, haah2018ProdDecPerFuncQSignPRoc}, this equation can be solved using a root finding method. Namely, we solve for the roots of the expression
\begin{align*}
1 - F(w) F(w^{-1})
\end{align*}
over \emph{all} complex numbers. Then, all real roots will come in pairs $\{r_i, 1/r_i\}_i$ while all non-real complex roots will come in quadruples $\{z_j, 1/z_j, \bar z_j, 1/ \bar z_j\}_j$. Hence, 
\begin{align*}
1 - F(w) F(w^{-1}) \propto \prod_i (w-r_i)(1/ w- r_i) \prod_j (w-z_j)(1/w - z_j) (w - \bar z_j) (1/w - \bar z_j).
\end{align*}
Evaluating this at $w=1$, the constant of proportionality has to be real and positive by the constraint on $F$. Denoting this by $\alpha$, we conclude 
$$G(w) = \sqrt{\alpha} \prod_i (w-r_i) \prod_j (w- z_j) (w- \bar z_j)$$
is a possible solution.\footnote{This is not unique since, for instance, we could have chosen the same with $w \mapsto 1/w$. } Note that by construction $G(w)$ is a real Laurent polynomial.

\paragraph{Decomposition via halving}
We now give an algorithm for the decomposition step in quantum signal processing motivated by Corollary~\ref{unitaryDecomp}. Let $U$ be a degree $d$ parity unitary in the Low algebra. We wish to find a unitary $V_1$ of degree $l$ respectively such that $V_1^* U$ is of degree at most $d-l$. In other words, we wish to solve for $V_1^*$ the coefficients of the $A, B$ Laurent polynomials in equation~\ref{exp1} such that the above is true. Denoting the coefficients respectively by $\{x_n\}_{n=-l}^{l}, \{y_n\}_{n=-l}^{l}$, we can express $V_1^*$ as
\begin{align*}
V_1^* = \sum_{n = -l}^{l} \begin{pmatrix}
    x_n & i y_n \\
    i y_{-n} & x_{-n} 
  \end{pmatrix} w^n.
\end{align*}
Note that the parity condition will eliminate half of the variables. The unitary $U$ can be put into a similar form:
\begin{align*}
U = \sum_{n = -d}^{d} \begin{pmatrix}
    a_n & i b_n \\
    i b_{-n} & a_{-n} 
  \end{pmatrix} w^n
\end{align*}
so that
\begin{align*}
  V_1^* U = \sum_{n = -d -l}^{d+l} \Bigg[ \sum_{\substack{\lvert n_1 \rvert \le l, \lvert n_2 \rvert \le d \\  n_1+ n_2 = n}}^{} 
  \begin{pmatrix}
    x_{n_1} & i y_{n_1} \\
    i y_{-n_1} & x_{-n_1} 
  \end{pmatrix} 
  \begin{pmatrix}
    a_{n_2} & i b_{n_2} \\
    i b_{-n_2} & a_{-n_2} 
  \end{pmatrix}
  \Bigg] w^{n}.
\end{align*}
Hence, we can see that the $A, B$ Laurent polynomials for $V_1^*U$ have degree $n$ term coefficients given by
\begin{align}
\label{VUCoef}
  \sum_{\substack{\lvert n_1 \rvert \le l, \lvert n_2 \rvert \le d \\  n_1+ n_2 = n}}^{}   a_{n_2}x_{n_1} - b_{-n_2} y_{n_1},
  \sum_{\substack{\lvert n_1 \rvert \le l, \lvert n_2 \rvert \le d \\  n_1+ n_2 = n}}^{} b_{n_2} x_{n_1} +a_{-n_2} y_{n_1},
\end{align}
respectively. Now, for the sake of brevity, assume $d,l$ are even. We can group all the coefficients of $V_1^*$ into a single $2(l+1)$ dimensional vector:
\begin{align*}
  \begin{pmatrix}
    x_{-l} \\
    x_{-l+2} \\
    \vdots\\
    x_l \\
    y_{-l} \\
    y_{-l+2} \\
    \vdots \\
    y_l
  \end{pmatrix}. 
\end{align*}
By equation~\ref{VUCoef}, we can show that right multiplication by $U$ is equivalent to a linear transformation given by the following $2(d+l+1) \times 2(l+1)$ dimensional matrix:
\begin{align*}
\scalebox{0.925}{$
  \begin{pmatrix}
  a_{-d}        &       0       &       0       & \cdots &       0       &       -b_d    &       0       &       0       &       \cdots   &       0 \\
  a_{-d+2}      &       a_{-d}  &       0       & \cdots &       0       &       -b_{d-2} &      -b_d    &       0       &       \cdots   &       0\\
  \vdots        &       \vdots  &       \vdots  &       \vdots  &       \vdots  &       \vdots  &       \vdots  &       \vdots  &       \vdots  &       \vdots\\
  a_{-d+2l}     &       a_{-d+2l -2}    & a_{-d+2l-4}   &       \cdots & a_{-d}  &       -b_{d-2l} &     -b_{d-2l+2}     &       -b_{d-2l+4}     &       \cdots   &       -b_d\\
  a_{-d+2l+2}   &       a_{-d+2l}       & a_{-d+2l-2}   &       \cdots & a_{-d+2}        &       -b_{d-2l-2} &   -b_{d-2l}       &       -b_{d-2l+2}     &       \cdots   &       -b_{d-2}\\
  \vdots        &       \vdots  &       \vdots  &       \vdots  &       \vdots  &       \vdots  &       \vdots  &       \vdots  &       \vdots  &       \vdots\\
  a_{d} &       a_{d-2} & a_{d-4}       &       \cdots & a_{d-2l}        &       -b_{-d} &       -b_{-d+2}       &       -b_{-d+4}       &       \cdots   &       -b_{-d+2l}\\
  0     &       a_{d}   & a_{d-2}       &       \cdots & a_{d-2l+2}      &       0 &     -b_{-d} &       -b_{-d+2}       &       \cdots   &       -b_{-d+2l-2}\\
  \vdots        &       \vdots  &       \vdots  &       \vdots  &       \vdots  &       \vdots  &       \vdots  &       \vdots  &       \vdots  &       \vdots\\
  0     &       0       &       \cdots   & 0     &       a_d     &       0       &       0       &       \cdots   &       0       &       -b_{-d} \\
b_{-d}  &       0       &       0       & \cdots &       0       &       a_d     &       0       &       0       &       \cdots   &       0 \\
  b_{-d+2}      &       b_{-d}  &       0       & \cdots &       0       &       a_{d-2} &       a_d     &       0       &       \cdots   &       0\\
  \vdots        &       \vdots  &       \vdots  &       \vdots  &       \vdots  &       \vdots  &       \vdots  &       \vdots  &       \vdots  &       \vdots\\
  b_{-d+2l}     &       b_{-d+2l -2}    & b_{-d+2l-4}   &       \cdots & b_{-d}  &       a_{d-2l} &      a_{d-2l+2}      &       a_{d-2l+4}      &       \cdots   &       a_d\\
  b_{-d+2l+2}   &       b_{-d+2l}       & b_{-d+2l-2}   &       \cdots & b_{-d+2}        &       a_{d-2l-2} &    a_{d-2l}        &       a_{d-2l+2}      &       \cdots   &       a_{d-2}\\
  \vdots        &       \vdots  &       \vdots  &       \vdots  &       \vdots  &       \vdots  &       \vdots  &       \vdots  &       \vdots  &       \vdots\\
  b_{d} &       b_{d-2} & b_{d-4}       &       \cdots & b_{d-2l}        &       a_{-d} &        a_{-d+2}        &       a_{-d+4}        &       \cdots   &       a_{-d+2l}\\
  0     &       b_{d}   & b_{d-2}       &       \cdots & b_{d-2l+2}      &       0 &     a_{-d}  &       a_{-d+2}        &       \cdots   &       a_{-d+2l-2}\\
  \vdots        &       \vdots  &       \vdots  &       \vdots  &       \vdots  &       \vdots  &       \vdots  &       \vdots  &       \vdots  &       \vdots\\
  0     &       0       &       \cdots   & 0     &       b_d     &       0       &       0       &       \cdots   &       0       &       a_{-d} 
  \end{pmatrix}.
  $
  }
\end{align*}
The structure of this matrix is that of a block banded Toeplitz matrix, which are studied for instance in~\cite{bini1988efficient}. Now, we can enforce the condition that $V_1^* U$ is of degree at most $d - l$ by extracting the rows for $n < -d +l$ and $n > d-  l$, resulting in the linear equations:
\begin{align}
  \label{theMatrix1}
\scalebox{0.85}{$
    \begin{pmatrix}
  a_{-d}        &       0       &       0       & \cdots        &       0       &       -b_d    &       0       &       0       &       \cdots  &       0 \\
  a_{-d+2}      &       a_{-d}  &       0       & \cdots &      0       &       -b_{d-2} &      -b_d    &       0       &       \cdots  &       0\\
  \vdots        &       \vdots  &       \vdots  &       \vdots  &       \vdots  &       \vdots  &       \vdots  &       \vdots  &       \vdots  &       \vdots\\
  a_{-d+2l-2}     &       a_{-d+2l -4}    & \cdots   &         a_{-d}  &   0   &      -b_{d-2l+2} &     -b_{d-2l+4}     &       \cdots     &       -b_d   & 0\\
  0     &       a_{d}   & a_{d-2}       &       \cdots &        a_{d-2l+2}      &       0 &     -b_{-d} &       -b_{-d+2}       &       \cdots  &       -b_{-d+2l-2}\\
  \vdots        &       \vdots  &       \vdots  &       \vdots  &       \vdots  &       \vdots  &       \vdots  &       \vdots  &       \vdots  &       \vdots\\
  0     &       0       &       \cdots  & 0     &       a_d     &       0       &       0       &       \cdots  &       0       &       -b_{-d} \\
b_{-d}  &       0       &       0       & \cdots        &       0       &       a_d     &       0       &       0       &       \cdots  &       0 \\
  b_{-d+2}      &       b_{-d}  &       0       & \cdots &      0       &       a_{d-2} &       a_d     &       0       &       \cdots  &       0\\
  \vdots        &       \vdots  &       \vdots  &       \vdots  &       \vdots  &       \vdots  &       \vdots  &       \vdots  &       \vdots  &       \vdots\\
  b_{-d+2l-2}     &       b_{-d+2l -4}    & \cdots   &        b_{-d}   &   0  &       a_{d-2l+2} &      a_{d-2l+4}      &       \cdots  &       a_d   &   0\\
  0     &       b_{d}   & b_{d-2}       &       \cdots &        b_{d-2l+2}      &       0 &     a_{-d}  &       a_{-d+2}        &       \cdots  &       a_{-d+2l-2}\\
  \vdots        &       \vdots  &       \vdots  &       \vdots  &       \vdots  &       \vdots  &       \vdots  &       \vdots  &       \vdots  &       \vdots\\
  0     &       0       &       \cdots  & 0     &       b_d     &       0       &       0       &       \cdots  &       0       &       a_{-d} 
  \end{pmatrix}
  \begin{pmatrix}
    x_{-l} \\
    x_{-l+2} \\
    \vdots\\
    x_l \\
    y_{-l} \\
    y_{-l+2} \\
    \vdots \\
    y_l
  \end{pmatrix} = 0,
  $}
\end{align}
where the matrix is of dimension $4l\times 2(l+1)$.
We also add the condition that $V_1^*(1) = I$, which amounts to additional linear equations
\begin{align}
    \sum_{n=-l}^l x_n = 1, \sum_{n=-l}^l y_n = 0,
    \label{theMatrix2}
\end{align}
resulting in a total of $4l+2$ linear equations.
By Corollary~\ref{unitaryDecomp} the system of linear equations given by equations~\ref{theMatrix1} and~\ref{theMatrix2} has a unique solution and returns a degree $l$ parity unitary $V_1^*$ such that $V_1^* U$ is of degree $d-l$. Note that this implies that the rows in the above matrices are not independent. 

\medskip

However, in real implementations the element we are to decompose might not be exactly unitary. This may be due to numerical imprecision or errors introduced in the completion step. We therefore propose to solve the above linear system by the least squares method. This gives a recursive algorithm:

\begin{framed}
\noindent\hypertarget{alg:decomp}{\textbf{Algorithm 1:} decompose(parity Low algebra element $M$)}

\vskip3mm

\begin{algorithmic}

\If{$\deg M = 1$}\State{return $\{M\}$}
\Else 
\State{solve equations~\ref{theMatrix1} and~\ref{theMatrix2} by least squares to get parity Low algebra element $M_1$}
\State{return $\mathrm{decompose}(M_1) \cup \mathrm{decompose}(M_1^* M)$}
\EndIf

\end{algorithmic}
\end{framed}

\paragraph{Runtime.} Since our algorithm involves root finding, in the worst case its runtime can be as high as cubic in degree $n$, similar to the algorithm of~\cite{haah2018ProdDecPerFuncQSignPRoc}. The halving procedure can also be performed in cubic time and potentially even faster~\cite{cohen2018LPSolving}.

\section{Experimental Results}
We perform numerical experiments implementing our algorithms and compare them to the algorithm given in~\cite{haah2018ProdDecPerFuncQSignPRoc}, which also used root finding for completion but implemented decomposition differently. Namely, given a degree $d$ unitary $U$ in the Low algebra, they sequentially solve for the linear terms $L_i = Q_i \wt w Q_i^{-1}$ in equation~\eqref{eq:unitaryDecomp} from $i = d$ to $1$. More specifically, they start with $U$, and solve for $L_d$. The same is then repeated for $L_d^* U$ to obtain $L_{d-1}$ and so on. We shall refer to this algorithm as \emph{carving}. Due to the iterative nature of this algorithm, the error introduced at the beginning would blow up during the carving process. Therefore, it requires very high precision arithmetic in the completion step. Our halving algorithm, on the other hand, can be performed with standard 64-bit machine precision throughout. 

\medskip

In the following, we report the results of experiments in which we tested our algorithm using the  Python \texttt{numpy} package on a MacBook Pro with 2.3 GHz Intel Core i5 processor and 16 GB memory.

\paragraph{Hamiltonian simulation.}
Hamiltonian simulation is one of the most important applications of quantum signal processing~\cite{childs2017towardsFirstQSimSpeedup,haah2018ProdDecPerFuncQSignPRoc,gilyen2018QSingValTransf, dong2020efficientPhaseFindingInQSP}. Formally, the problem is the following. Given a Hermitian $H$, with $\|H\|\le1$, and time-interval for the evolution $\tau$, we wish to implement the unitary $e^{-iH\tau}$. When a quantum walk is used~\cite{childs2008OnRelContDiscQuantWalk,berry2012BlackHamSimUnitImp}, the unitary $W$ can be implemented with a quantum circuit, whose eigenvalues $\mp e^{\pm i\theta_\lambda}$ are associated with the eigenvalues $\lambda$ of the Hamiltonian via 
$$\sin2\theta_\lambda = \lambda.$$
The task is then to achieve the effect of the function $F:e^{i\theta}\rightarrow e^{i\tau\sin{2\theta}}$. More specifically,
\begin{equation}\label{bessel}
F(w)=\exp\left(\tau\frac{w^2-w^{-2}}{2}\right)=\sum_{k\in\mathbb{Z}}J_k(\tau)w^{2k},
\end{equation}
where $J_k$ are the Bessel functions of the first kind. In Equation (\ref{bessel}), the function $F(w)$ is already in the form of a Laurent series whose coefficients decrease exponentially. One can approximate $F(w)$ up to an additive error $\epsilon$ for $w\in U(1)$ by truncating it into degree $n=2\lceil \frac{e}{2}\tau+\ln(1/\epsilon)\rceil$~\cite{haah2018ProdDecPerFuncQSignPRoc},
\[\hat{F}_\epsilon(w)=\sum_{k=-n/2}^{n/2}J_k(\tau)w^{2k}.\] Empirical results also suggest that one needs to scale down $\hat{F_\epsilon}(w)$ by a factor $\eta\in(0,1)$, in order to have more numerical stability during the root finding completion. However, the downscaling would decrease the success  probability in the post-selection by a factor of $\eta$. This can be regarded as a tradeoff between the classical preprocessing phase versus the actual time complexity of the algorithm. 

\medskip

Specifically, the inputs to the algorithm are the time-interval $\tau$, error tolerance $\epsilon$ and scaling factor $\eta$. We then further truncate the list of coefficients of $\eta\cdot\hat{F}_{\epsilon}(w)$ to standard double precision. The resulting Laurent polynomial would be the input of the angle finding algorithm.

\medskip

\begin{figure*}[ht]
\centering
 \includegraphics[width=.8\linewidth]{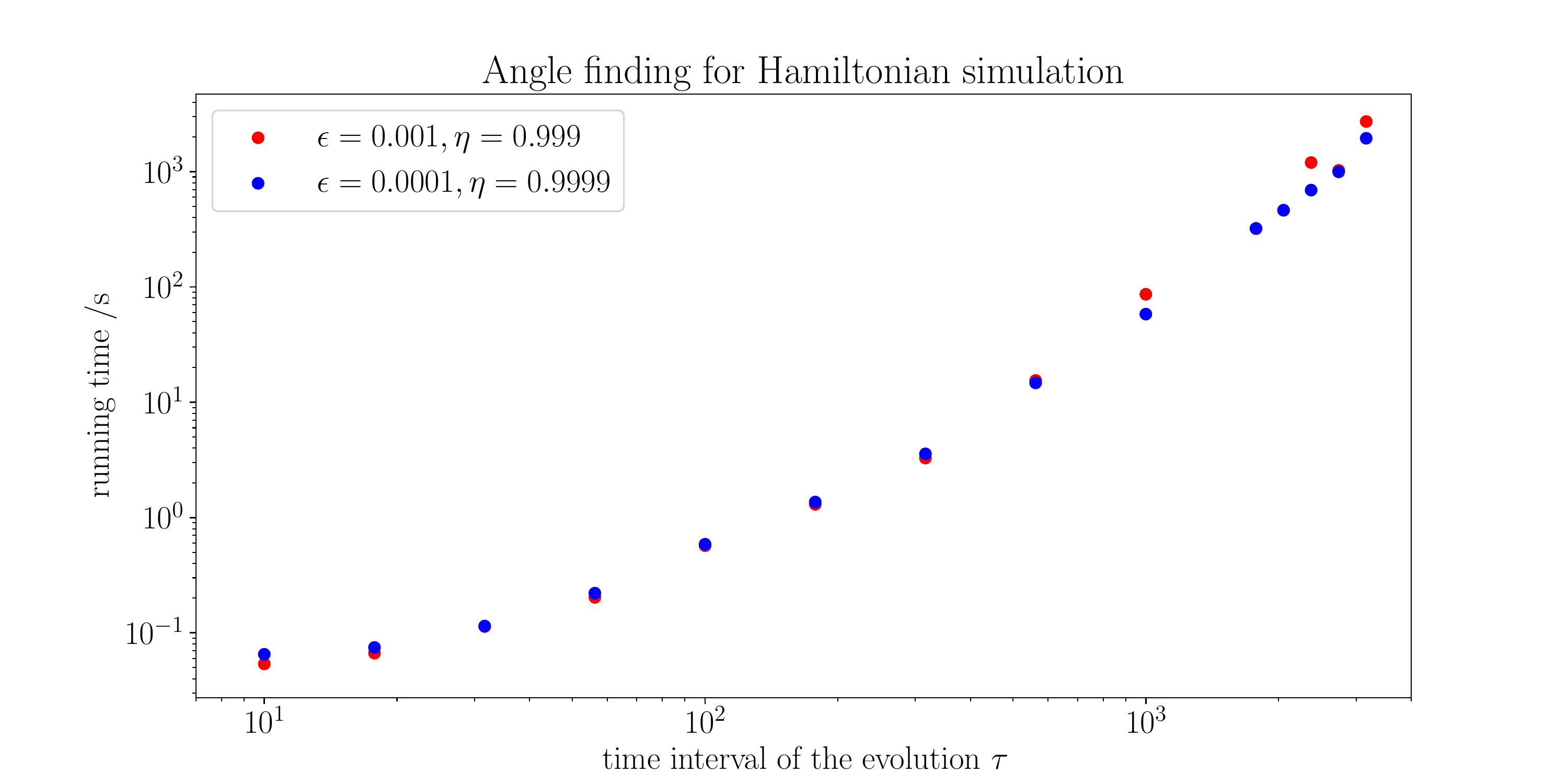}
\caption{The running time for angle finding using the halving algorithm. Here we show the results for $\epsilon  = 10^{-3}, \, \eta = 0.999$ and $\epsilon = 10^{-4}, \, \eta = 0.9999$. The running time scales as a cubic function with respect to the degree of the Laurent polynomial and hence also cubic with respect to the evolution time parameter $\tau$. Note that an instance with $\tau=1200$, corresponding to a Laurent polynomial with degree 3261, can be efficiently solved within $5$ minutes.}
\label{fig:time}
\end{figure*}

We performed angle decomposition for different $\tau$, with fixed error tolerance $\epsilon=0.001$ and $\eta = 0.999$. For the experiments, the Laurent polynomial is truncated to guarantee an error upper bound of  $0.1\epsilon$, with the capitalizing parameter, i.e.\ the coefficients of the appended highest- and lowest-degree terms, set to $0.45\epsilon$. The detailed numerical results are illustrated in Fig.~\ref{fig:time}.

\paragraph{Comparison to the carving algorithm.} We compare the precision requirements demanded by the halving and carving algorithms with respect to the time-interval of the evolution $\tau$ and the error tolerance $\epsilon$. Note that computing the sequence of angles becomes more difficult with larger $\tau$ and smaller $\epsilon$. We call a parameter pair $(\tau, \epsilon)$ \emph{achievable} with respect to an angle finding algorithm, if that algorithm can output a sequence of angles with the effect of a Laurent polynomial $\epsilon$-close to the Hamiltonian simulation function $e^{\tau\frac{w^2-w^{-2}}{2}}$ with only machine precision. The Laurent polynomial is truncated to guarantee an error upper bound  $\epsilon/10$, with capitalizing parameter set to $\epsilon/3$.

\medskip

\begin{figure*}[h]
\centering
 \includegraphics[width=.6\linewidth]{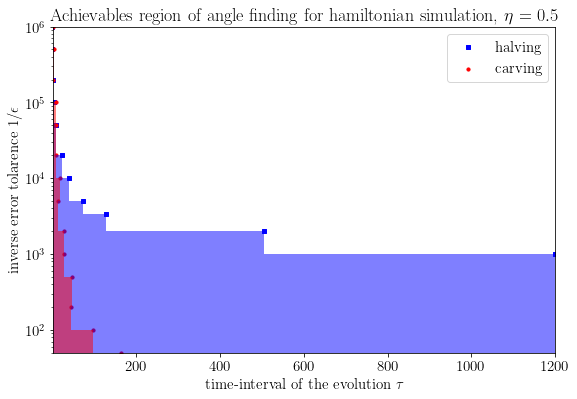}
\caption{Comparison between the halving and carving algorithms' achievable parameter regions for the Hamiltonian simulation problem with machine precision. Note that the $y$-axis is log scaled.}
\label{fig:compare}
\end{figure*}

We can determine the achievable regions through numerical means and plot them in Fig.~\ref{fig:compare}. It can be seen that the achievable region for the halving algorithm is far larger than the carving algorithm. 

\section{Further Remarks}
\paragraph{Empirical error analysis.} In \cite{haah2018ProdDecPerFuncQSignPRoc}, an error analysis is carried out to show that the precision $p_c$ needed for carving scales as $poly(n,\log(1/\epsilon))$ in order to solve for angle sequences for a Laurent polynomial with degree $n$ and error tolerance $\epsilon$. When the leading coefficients decay exponentially, the carving algorithm would suffer from numerical instability, although this issue can be easily resolved by setting each coefficient to be a multiple of $\epsilon/n$. The halving algorithm had the same issue, but this was resolved by capitalization.

\medskip
Denoting the magnitude of the largest leading coefficients by $\delta$, it is plausible that for carving, $$p_c=poly(n,\log(1/\epsilon), \log(1/\delta)).$$

Our experimental results give evidence that the precision $p_h$ needed for halving scales as $$p_h=poly(\log n, \log(1/\epsilon),\log(1/\delta)),$$ since it does not suffer from the error blow-up from the iterative carving. If so, then the halving method would be a truly numerically stable algorithm. We experimentally observe that this is almost certainly the case for a particular family of random inputs, where each of the coefficients of the input polynomial is drawn i.i.d.\ under a normal distribution.

\medskip

However, for small degrees, a family of pathological Low algebra elements can be found where the final error scales as $\delta^{-n/2}$ under machine precision. This phenomenon can be remedied by carefully re-choosing root pairs in the completion step. It is observed that the final error scales as $n/\delta$ after implementing this ad hoc solution. It is not clear whether there is an efficient way to choose roots such that the algorithm is provably numerically stable. In the current implementation, two complex roots out of four, or one real root out of two, are randomly chosen. This is done so that the coefficients of the Laurent polynomial we obtain does not grow or decay exponentially, as pointed out in~\cite{haah2018ProdDecPerFuncQSignPRoc}. Ideally this should be derandomized and also devised to guarantee numerical stability in the decomposition step. We leave this for future work.

\paragraph{Integration with optimization-based angle sequence finding.} In~\cite{dong2020efficientPhaseFindingInQSP}, a novel algorithm is proposed to solve for the angle sequence via an optimization approach. This algorithm makes completion unnecessary and thereby sidesteps root finding, a subroutine that can be time-consuming and prone to numerical error. Starting from carefully chosen initial points, experiments show that the optimization-based algorithm finds the angle sequence with a higher precision than approaches based on root finding. However, the major open problem of~\cite{dong2020efficientPhaseFindingInQSP} is to prove that the algorithm always finds the global optimum starting from the initial points. This can be difficult considering  that they observe an extremely complex landscape for the objective function they wish to optimize. 

\medskip

It could be possible to create a hybrid algorithm combining  halving and black-box optimization: A good initial point for the angle sequences is found via the halving algorithm and is then fed into the black-box optimizer to further boost the precision. The number of queries can be upper bounded by analyzing the landscape of the objective function in the vicinity of the optimum point, which could in principle be easier than analyzing the global landscape. We leave this to future work.

\section*{Acknowledgments} We thank Andrew Childs, Yuan Su and Yaoyun Shi for useful discussions. DD and AG would like to thank God for all of His provisions.

\newpage


\bibliographystyle{alphaUrlePrint}
\bibliography{Bibliography,LocalBibliography}

\end{document}